# Air pollution forecasting by coupled atmosphere-fire model WRF and SFIRE with WRF-Chem

Adam K. Kochanski[A], Jonathan D. Beezley[B], Jan Mandel[C], and Craig B. Clements[D]

[A] Department of Atmospheric Sciences, University of Utah, UT, USA, adam.kochanski@utah.edu
[B] Météo France and CERFACS, Toulouse, France, jon.beezley@gmail.com
[C] Department of Mathematical and Statistical Sciences, University of Colorado Denver, Denver, CO, USA, jan.mandel@ucdenver.edu
[D] Department of Meteorology and Climate Science, San José State University, San José, CA, craig.clements@sjsu.edu

**Abstract:**

Atmospheric pollution regulations have emerged as a dominant obstacle to prescribed burns. Thus, forecasting the pollution caused by wildland fires has acquired high importance. WRF and SFIRE model wildland fire spread in a two-way interaction with the atmosphere. The surface heat flux from the fire causes strong updrafts, which in turn change the winds and affect the fire spread. Fire emissions, estimated from the burning organic matter, are inserted in every time step into WRF-Chem tracers at the lowest atmospheric layer. The buoyancy caused by the fire then naturally simulates plume dynamics, and the chemical transport in WRF-Chem provides a forecast of the pollution spread. We discuss the choice of wood burning models and compatible chemical transport models in WRF-Chem, and demonstrate the results on case studies.

## 1. Introduction

The adverse effects of smoke on air quality and visibility are of great concern for fire and land managers planning prescribed burns. The Federal Wildland Fire policy and the Clear Air Act significantly broaden regulatory and management requirements by demanding from fire and land managers an assessment of the air quality and visibility impacts from wildfires and fire management programs. Since the fire emissions may lead to a violation of the National ambient air quality standards, decisions about the prescribed burns must be made not only based on the fire safety criteria (wind speed, fuel moisture etc.) but also taking into account impacts of the smoke on air quality and visibility. In this paper, we describe a newly added functionality of the coupled atmosphere-fire model WRF-SFIRE, allowing for the simulation and forecasting of the smoke emissions, dispersion, and their effect on the air quality both on the local and regional scale.

Smoke forecasting is a very complex problem that spans across many disciplines. In order to estimate smoke spread and its effects on air quality, first one has to assess the amount of the fuel burnt, and convert it into emission fluxes of the particular chemical species (or smoke in general). In the next step, the vertical distribution of the smoke in the atmosphere (injection height and vertical profile) has to be estimated, based on the fire intensity and meteorological conditions (winds, atmospheric stability). Knowing the smoke injection characteristics and local flow pattern, allows for an estimation of the smoke dispersion and deposition. In the last step, the chemical processes associated with smoke dispersion in the atmosphere have to be taken into account in order to assess the smoke impact on the air quality. As described above, the smoke dispersion is clearly a multidisciplinary problem, with one common denominator though – the weather, which effects fuel characteristics, fire behavior (intensity and amount of fuel burnt), smoke injection, smoke dispersion and smoke chemistry. From this standpoint, building a smoke forecasting system around the weather forecasting system seems to be a logical choice, and in this paper we present a first attempt toward creating such a model.

There is a wide suite of tools of various complexity levels that may be helpful in assessing the smoke dispersion. They range from simple Gaussian smoke models like VSMOKE (Lavdas 1996) and SASEM (Sestak and Riebau 1988) providing the area affected by smoke based on the defined location, fuel type, fire area and wind conditions like, through puff models like CALPUFF (Scire 2000) to complex multi-model systems like BlueSky (Larkin et al. 2009) providing estimation of fire emissions, dispersion and air quality effects associated with fires. For more information on smoke transport models please see the smoke modeling review by Goodrick et al. (2012). Since the weather plays a key role in fire progression and dispersion of the fire emissions, some sort of a weather forecast is generally required for any of these tools in order to make any prediction how the fire smoke will disperse and affect the air quality. In the simplest case, it may be in a from of a user input who must define the wind speed and direction for which the smoke dispersion will be assessed. In complex systems like BlueFire, it may come from a separate numerical weather forecasting model providing the weather input to the fire emission and dispersion components of the system. Note that in the complex modeling framework like BlueSky, the weather input generally affects a number of its components. For instance, the meteorological conditions can be taken into account for the estimation of the burn rates, but also may be used in the plume rise model computing the vertical smoke distribution as well the dispersion and chemical models like CMAQ. WRF-SFIRE bases on a similar principle, but all the components are integrated around the weather forecasting system. The model comprehensively resolves the fire progression, the heat release associated with the fire, its plume rise and smoke dispersion and chemistry without any external components, providing the weather, fire progression, emission and air quality forecasts.

## 2. Model description

The core of the system is the WRF-SFIRE model, which is a two-way coupled fire atmosphere model based on WRF (Skamarock et al. 2005). It provides forecast of the fire spread based on the local meteorological conditions, taking into account the feedback between the fire and the atmosphere (Mandel et al. 2011). In order to capture the effect of local weather on fuel characteristics, WRF-SFIRE is also coupled with a fuel moisture model. That allows the model to forecast the fuel moisture based on the local meteorology (Kochanski et al. 2012). The fire model is also coupled with WRF-Chem so the smoke emitted from the fire is transported within the atmosphere, and undergoes chemical reactions resolved by WRF-Chem. The diagram showing the WRF-SFIRE model components is presented in Figure 1.

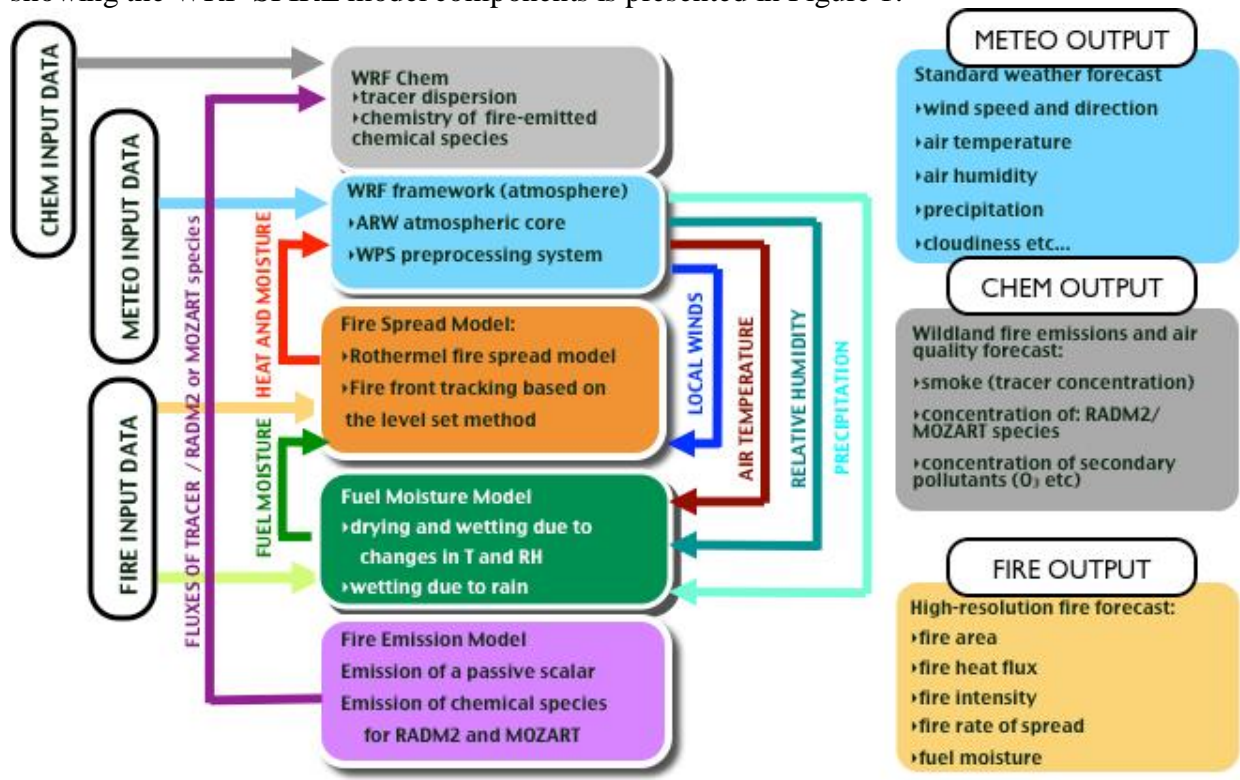

Figure 1. Diagram of WRF-SFIRE coupled with moisture model and WRF-Chem

In each model time step, the near-surface wind from WRF is interpolated vertically to a logarithmic profile and horizontally to the fire mesh to obtain height-specific wind that is input into Rothermel fire spread-rate formula (Rothermel 1972). Fuels are given as one of 13 categories (Anderson 1982), and associated with each category are prescribed fuel properties such fuel mass, depth, density, surface-to-volume ratio, moisture of extinction, and mineral content. The fuel moisture content is computed based on relative humidity of the air, air temperature and precipitation forecasted by the atmospheric component of WRF. Based on these fuel properties and WRF-SFIRE winds, the instantaneous fire spread rate at every refined mesh point is computed. After ignition, the amount of fuel remaining is assumed to decrease exponentially with time with time, with the time constant dependent on fuel properties, and the latent and sensible heat fluxes from the fuel burned are inserted into the lowest levels of the atmospheric model, assuming exponential decay of the heat flux with height. The full description of the WRF-SFIRE can be found in Mandel et al. 2011. The current code and documentation are available from OpenWFM.org. A version from 2010 is distributed with the WRF release as WRF-Fire (Coen et al. 2012; OpenWFM 2012).

There are two mechanisms by which the smoke emissions are treated in the model. In the most complete form, the fire-emitted smoke is treated as sum fluxes of WRF-Chem compatible chemical species found. Fluxes of each chemical species are computed separately and ingested into the first layer of the WRF-Chem. The WRF-SFIRE currently uses FINN global emission factors (Wiedinmyer et al. 2011), which are based on the MODIS land cover types. Therefore, the first step in WRF-SFIRE is the conversion of the standard fuel categories (Anderson 1982), to the MODIS Land Cover Types (see Figure 2). After this conversion, the fuel consumption rates are computed for each fire grid point based on the mas of fuel burnt within one time step. In the next step the emission fluxes are computed as the products of the consumption rates and the fuel-specific emission factors. The FINN emission factors are directly compatible with the Model for Ozone and Related chemical Tracers (MOZART, Emmons 2010), so if that is the chemical scheme chosen by a user, the fluxes computed as described above are fed directly into the WRF-Chem. If other chemical scheme is to be used within WRF-Chem, the MOZART-compatible chemical fluxes have to be converted to the set of species compatible with that chemical scheme. As for now, WRF-SFIRE supports MOZART (natively) and RADM2 (trough the remapping of chemical species), as described by Emmons et al. (2010).

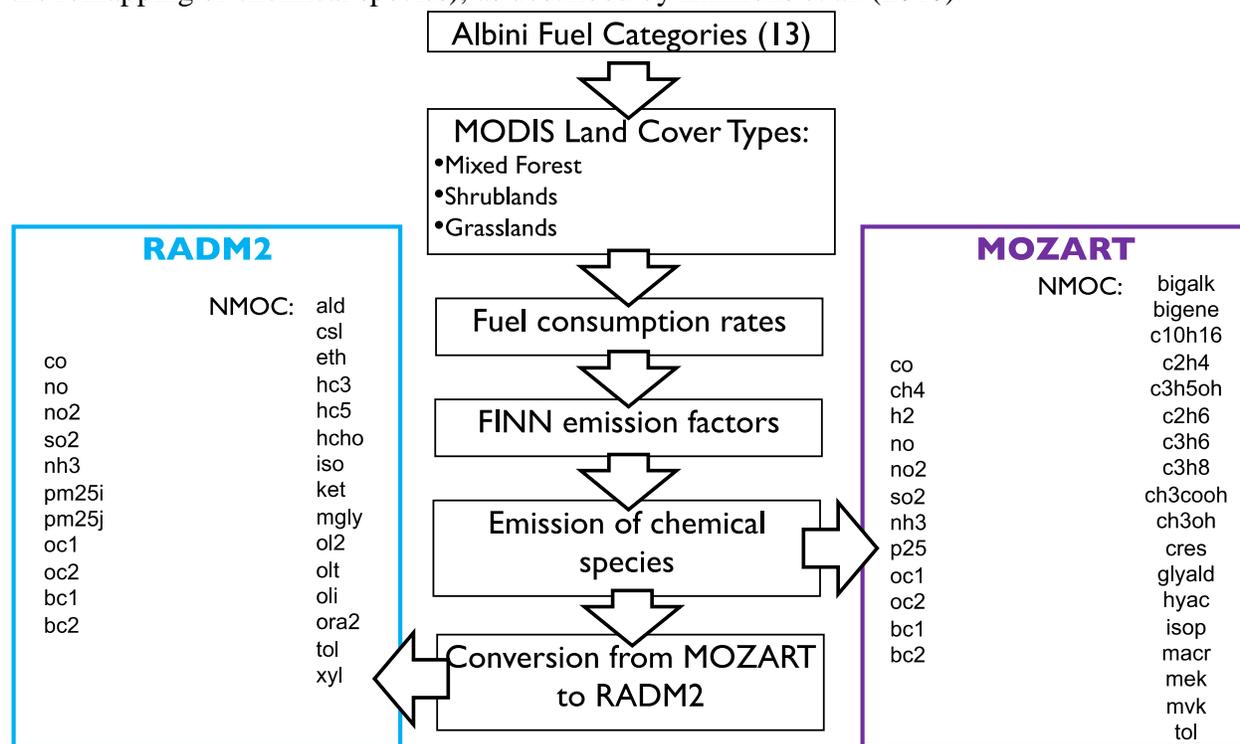

Figure 2. Computation of the smoke emissions in WRF-SFIRE coupled with WRF-Chem.

Treatment of the smoke as a mixture of the chemically active species that undergo chemical and physical processes in the atmosphere is very computationally intensive, since each additional chemical species leads to an additional 3D scalar equation, plus the complex mathematical representation of the chemical reactions. In order to reduce this computational cost, WRF-SFIRE offers also a simplified smoke representation trough a passive scalar. In this case, instead of resolving the concentration of all the spices listed in Figure 2, only one variable is added to the WRF computations (tr_8), which is a scalar that does not react in the atmosphere. If the simplified version of the smoke treatment is chosen, the execution of WRF-Chem is bypassed and the dynamical WRF core handles the transport of the smoke directly. In this simplified mode, the tracer flux is proportional to the fuel consumption rate.

## 3. Model setup

The two Santa Ana fires (2007) simulated in this study were driven by strong westerly Santa Ana winds induced by a high-pressure system located over northern Nevada. As the pressure built up and the high pressure system moved eastward, very strong and gusty Santa Ana winds bringing very warm and dry air from the Nevada desert affected the San Diego area. Wildland fires are directly driven by the local winds, often of a different characteristic than the main synoptic flow. Regional topography and land use mosaic may interact with the large-scale flow, creating specific local weather conditions that may be crucial for wildfire behavior. In order to resolve the development and movement of this large-scale weather system driving the Santa Ana winds, together with the local circulation affected strongly by the complex topography of southern California, WRF was configured with four nested domains: D01, D02, D03, and D04, of horizontal-grid sizes 32km, 8km, 2km, and 500m, respectively. The domain setup used in this study is shown in Figure 3. The fire model uses 30m-resolution elevation and fuel dataset, while the atmospheric model ~1.5km resolution MODIS land use representation.

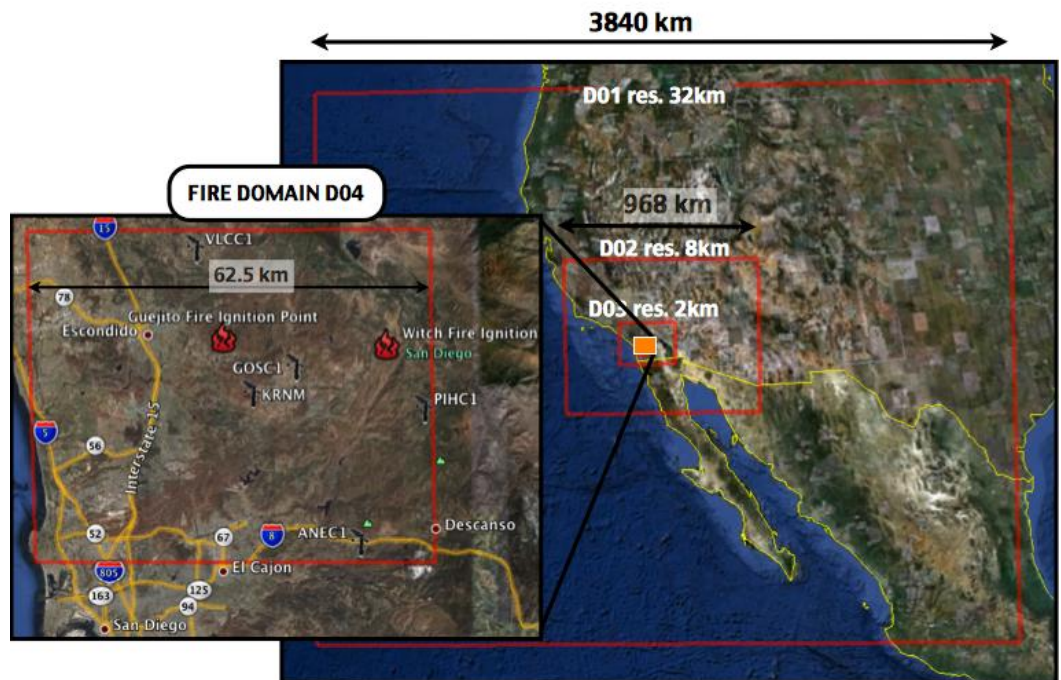

Figure 3. The multi-scale WRF setup in this study, with locations of fire origins and local meteorological stations used for model validation. Horizontal domain resolutions vary from 32km (D01) to 500m (D04).

The model simulation, initialized and driven by the North American Regional Reanalysis (Mesinger et al. 2006), was run for a period of 48h starting on 10.21.2007 5:00 am local time (12:00 UTC). The two simulated fires were initialized as point ignitions, and allowed to spread freely across the fuel mosaic driven by the local winds. The fire-emitted heat and moisture were feed into the atmospheric grid in order to mimic the fire-atmosphere coupling. The spatially-variable fuel moisture was initialized based on the station observations and kept constant through the whole simulation. The full description of the model configuration and the datasets used for model initialization may be found in Kochanski et al. 2013.

## 4. Results from the simulation of 2007 Santa Ana fires

The witch fire was ignited at 12:15 pm local time, roughly 7 h since the beginning of the simulation. The very strong Santa Ana wind blowing from ENE at a speed of up to 19 m/s (42.5 mph) begun to rapidly advance the fire toward. Due to the strength of the wind and not much variation in the simulated wind direction, the smoke was initially confined to a long but relatively narrow area located downwind from the ignition area, and pointing toward Encinitas (see Figure 4 a).
As the fire progressed and expanded on its flanks the span-wise extent of the plume increased as shown in Figure 4 b. Since the model estimates smoke emission based on the fuel consumption rate (fire intensity), smoke emission within the fire perimeter is not uniform. There is very intense plume emitted at the fire head and its flanks, while inside of the fire perimeter, where the amount of available fuel is significantly depleted, the smoke emission is less intense. The panel b shows smoke prediction 6 hours after ignition (5 h after the snapshot in panel a was taken). It is noteworthy how different the area affected by smoke is as compared to the previous time snapshot. Slight change in the wind direction in the western half of the domain pushed the smoke significantly further south than could be expected from the initial plot presented in panel a. Further variations in the wind direction led to gradually increasing spanwise extent of the plume which at that moment covers a big part of south-western corner of the simulation domain. Note the sharp south-eastern edge of the plume where steep canyons channel the flow limiting the smoke dispersion. On October 22$^{nd}$ at 1:00 am Guejito fire started at the Guejito Creek drainage, on the South Side of California State Route 78. Figure 4 c) shows the fire perimeters and smoke propagation two hours after the ignition of the Guejito fire. Note that at that moment, these are two separate fires emitting smoke independently, and even though the Guejito fire is much smaller than the Witch fire, its contribution to the overall smoke emission is clearly visible in the southern part of Escondido. As both fires expand, and approach each other, their plumes combine (Figure 4 d), before the actual fire perimeters merge. Note the non-homogeneous smoke emission at that moment. As the Witch fire approaches Ramona, its western edge becomes relatively inactive, which results in a visible smoke depletion in this region. At that moment, most of the smoke seems to be emitted by the northern and southern flanks of the Witch fire with significant contribution from the Guejito fire. Panel f) presents situation after both fires merged. The former Guejito fire extends significantly southward and becomes a strong source of fire emissions affecting Poway area. There are still active fire regions (hot spots) within the fire perimeter, as well as on the eastern edge of the Witch fire, which contribute to the overall smoke emissions. The fire shape and smoke at the end of the simulation is shown in Figure 4 f). This figure shows very complex fire and smoke emission pattern. There are multiple hot spots located on the northwestern fire perimeter, as well as in the center of the burnt area where slowly burning fuel is still being consumed by the fire. There are also active fire regions located at the leading edge of the northwestern and southwestern fire front where the fire advances downwind.

a) 10.21.2007 13:15 local time 1h after ignition of the Witch fire

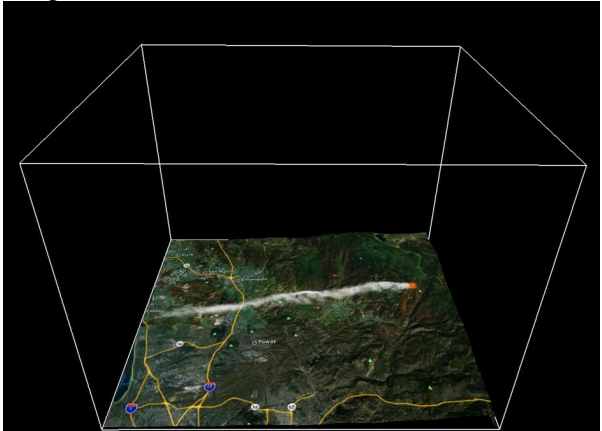

b) 10.21.2007 18:15 local time 6h after ignition of the Witch fire

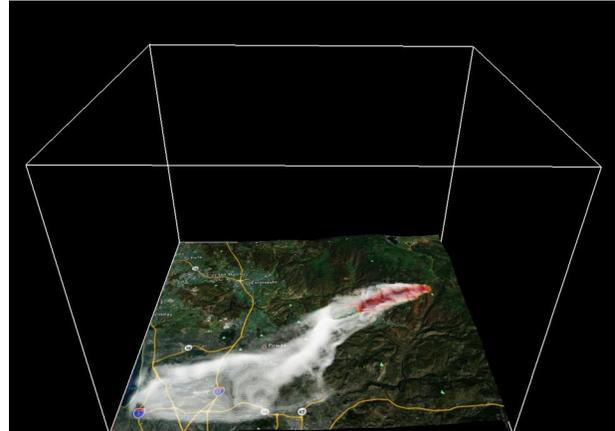

c) 10.22.2007 03:00 local time 2h after ignition of the Guejito fire

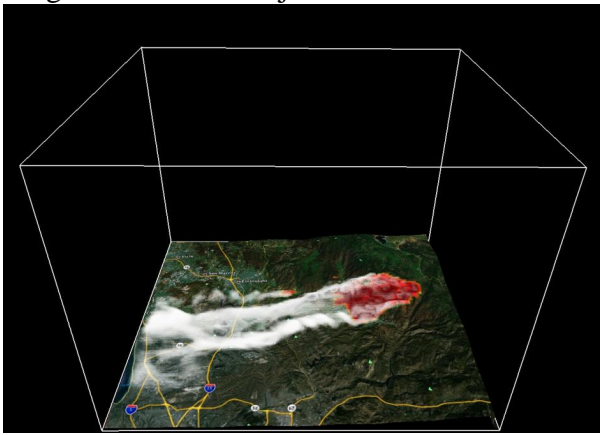

d) 10.22.2007 06:00 local time 5h after ignition of the Guejito fire

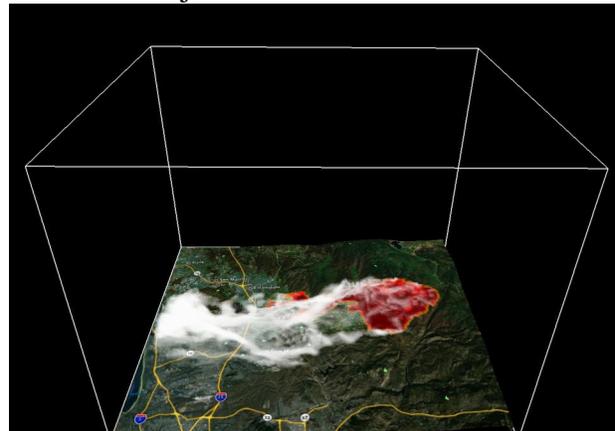

e) 10.22.2007 17:00 local time 36h into simulation

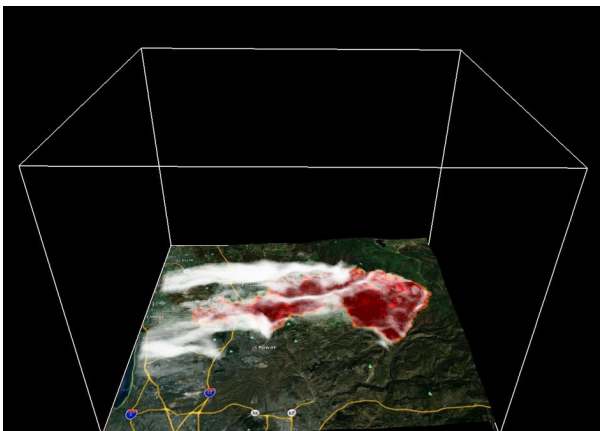

f) 10.23.2007 05:00 local time 48h into simulation

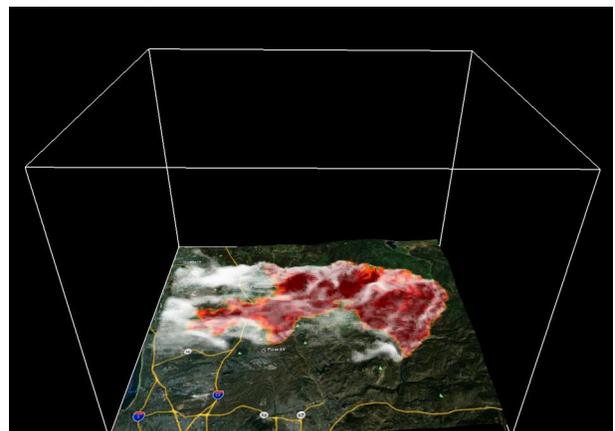

Figure 4. Visualization smoke (white) from the Witch and Guejito fires simulated in domain d04. Dark red fill represents totally burnt area (no fuel left), orange show active fire regions.

The nesting capabilities of WRF (used in this study) allow for running the model in multi-scale configurations, where the outer domain, set at relatively low resolution, resolves the large-scale synoptic flow, while the gradually increasing resolution of the inner domains allows for realistic representation of smaller and smaller scales, required for realistic rendering of the fire behavior and smoke emission. The two-way coupling between the domains allows for a feedback between the inner and outer domains.

a) WRF domain d03 (2km res) for 10.23.2007    b) MODIS image for 10.23.2007

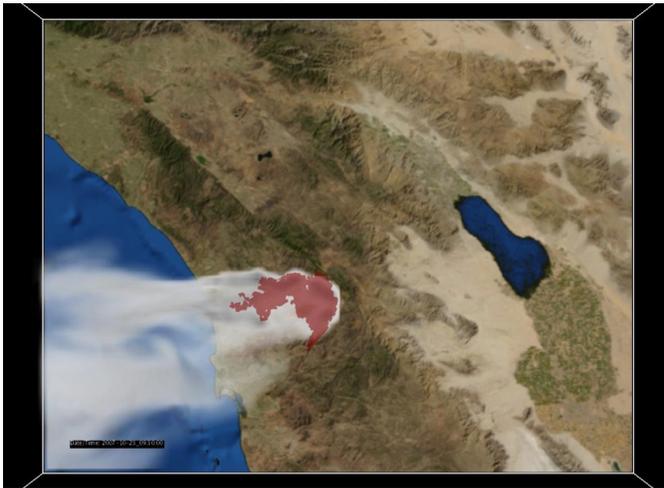 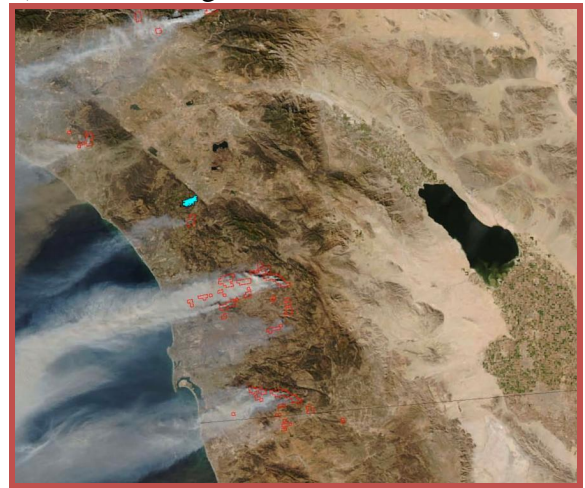

Figure 5. Smoke dispersion within domain d03 (2km resolution) simulated by WRF a) and MODIS satellite image b). The red color fill in panel a) represents the fire area projected from the nested fire domain d04 (500m resolution), red contours in panel b) represent remotely detected hot spots (regions of the highest fire intensity).

This mechanisms is used for the smoke transport from the innermost, high-resolution fire domain, to the coarser outer domains that can resolve large scale smoke transport. An example of the smoke dispersion within the coarser domain d03 is shown in Figure 5 a). Panel b) presents corresponding satellite image from MODIS (The Moderate Resolution Imaging Spectroradiometer). It should be noted that the MODIS image presented on the right panel represents an overall picture being an overall result of the smoke emitted from all fires in the region, while the WRF simulation shows only the effects of the Witch and Guejito fires. Nonetheless, there is a visible resemblance between the simulated and observed smoke dispersion. The aerial extend of the smoke is similar and even some of the smoke dispersion features off the coast of San Diego are visible in the WRF simulation. It is noteworthy that the location of the hot spots in the MODIS image (Figure 5b) corresponds well to the simulated active fire regions presented in Figure 4 a).

The results presented above come from the simplified WRF simulation in which smoke was represented just a as a passive tracer not reacting chemically in the atmosphere. Results from the same case but run with the full chemical smoke representation are shown below.

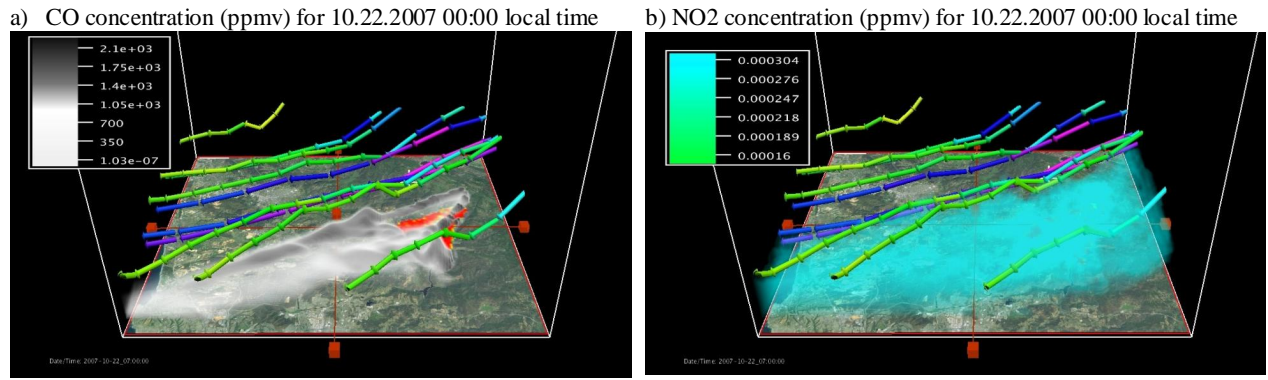

Figure 6. Fire emissions from domain (d04) simulated by WRF-SFIRE coupled with WRF-Chem. Color stream lines show the snapshot of the flow on 10.22.2007 at 00:00 local time. Orange file beneath the left panel represents fire area.

The chemical emissions for all chemical species and aerosols are estimated similarly, based on the fuel type, and the fuel consumption rate dependent on the fire. Therefore the concentration plots for all chemical species in the fire area are very similar. As an example of the model output, we have selected the concentration plots of carbon monoxide (CO) and nitrous dioxide ($NO_2$), which are shown in Figure 6 a) and b) respectively.

Figure 7 shows the streamlines overlaid over the ozone concentration simulated for the outer domain d03 (2km resolution). The streamlines indicate strong downslope Santa Ana wind pushing the fire to the coast. The elevated ozone concentrations of the in the wake of the Witch and Guejito fires suggest an impact of the fire emissions on the formation of the tropospheric ozone.

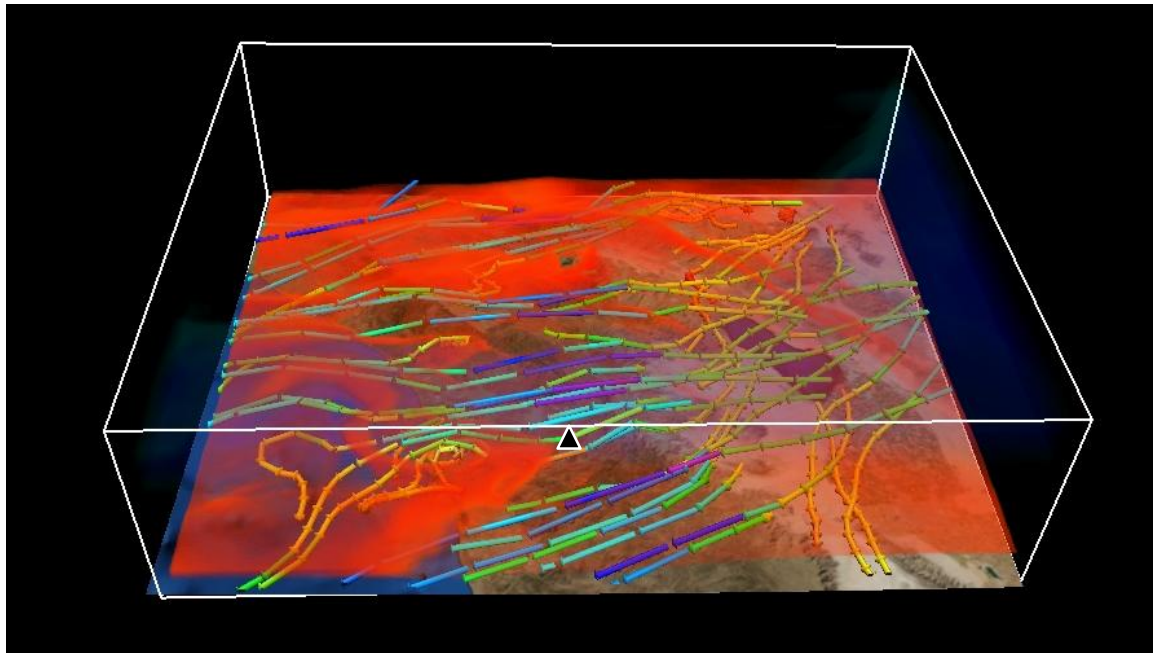

Figure 7. Elevated ozone concentration in the wake of the Witch and Guejito fires simulated in domain d03. The black triangle shows approximate location of the Witch fire ignition.

# 5. Conclusion

We have presented a scheme for coupling the emissions from a coupled atmosphere-fire model WRF-SFIRE into the atmospheric chemistry and transport model WRF-Chem. In this study, we showed results from the very first numerical simulations performed by a two-way coupled system of fire and weather with output to a chemical transport model, allowing for simulating the fire spread, smoke emission and dispersion, as well as smoke chemical transformations. Since it is a first attempt toward creating such a system, and the smoke emission component has not been validated yet, the results shown at this stage should be treated as a proof of concept, highlighting only general capabilities of WRF-SFIRE coupled with WRF-Chem.

We believe that the new capabilities added to WRF-SFIRE significantly increase its potential as a tool for future use by fire and land managers. Now, the fire spread forecast may be directly linked with the smoke emission, transport, and chemical conversion. As a result, besides the weather and fire spread forecast, the model can provide also smoke and air quality forecasts allowing for an estimation of comprehensive effects of prescribed burns. We also hope that the high levels of detail, provided by this system, will result in an improvement of the smoke emission and dispersion forecast as the model renders regions of the highest fire intensity, smoke emission and plume injection heights.

The high level coupling between the model components gives an opportunity for studying complex interactions between the fire and the atmosphere, including radiative and microphysical effects of the aerosols ingested into the atmosphere by fire plumes.

# Acknowledgements

This research was partially supported by the National Science Foundation (NSF) grants AGS- 0835579 and DMS-1216481, and National Aeronautics and Space Administration (NASA) grants NNX12AQ85G and NNX13AH9G. This work partially utilized the Janus supercomputer, supported by the NSF grant CNS-0821794, the University of Colorado Boulder, University of Colorado Denver, and National Center for Atmospheric Research.

Skamarock WC, Klemp JB, Dudhia J, Gill DO, Barker DM, Wang W, Powers JG (2005) A description of the advanced research WRF version 2. National Center for Atmospheric Research, Technical Note, NCAR/TN-468þSTR. (Boulder, CO)

Wiedinmyer C, Akagi SK, Yokelson RJ, Emmons LK, Al-Saadi J A, Orlando JJ, and Soja AJ (2011) The Fire INventory from NCAR (FINN): a high resolution global model to estimate the emissions from open burning, *Geoscientific Model Development (GMD)*, 4, 625-641, [doi:10.5194/gmd-4-625-2011](doi:10.5194/gmd-4-625-2011), 2011.